\documentclass{kapproc} 

\usepackage{procps} 

\usepackage[dvips]{graphicx}

\upperandlowercase

\kluwerbib 

\begin{document}

\articletitle{Chemical Evolution of Late-type Dwarf Galaxies}

\articlesubtitle{The windy starburst dwarfs NGC\,1569 and NGC\,1705}

\author{Donatella Romano,\altaffilmark{1} Monica Tosi,\altaffilmark{1} and 
        Francesca Matteucci\altaffilmark{2}}

\altaffiltext{1}{INAF--Osservatorio Astronomico di Bologna\\
                 Via Ranzani 1, 40127 Bologna, Italy}
                \email{[donatella.romano; monica.tosi]@bo.astro.it}

\altaffiltext{2}{Dipartimento di Astronomia, Università di Trieste\\
                 Via G.B. Tiepolo 11, 34131 Trieste, Italy}
                \email{matteucci@ts.astro.it}

\begin{abstract}
Thanks to the capabilities of modern telescopes and instrumentation, it is now 
possible to resolve single stars in external dwarf galaxies, provided they are 
bright enough. For galactic regions with deep enough photometry, detailed 
colour-magnitude diagrams are constructed, from which the star formation 
history and the initial mass function can be inferred by comparison with 
synthetic diagrams. Both the star formation history and the initial mass 
function are free parameters of galactic chemical evolution models. In this 
contribution we show how constraining them through high resolution photometry 
in principle allows us to better understand the mechanisms of dwarf galaxy 
formation and evolution.
\end{abstract}


\section{Introduction}

Low-luminosity galaxies -- dwarf galaxies and related systems -- are found 
numerous in the nearby universe. Here we deal with the subclasses of dwarf 
irregulars (DIGs) and blue compact dwarfs (BCDs). Both have low mass, low 
metallicity, large gas content and mostly young stellar populations. However, 
while BCDs are rather compact objects, with centrally concentrated starburst, 
gas and star distributions, DIGs are dominated by scattered bright H\,{\small 
II} regions in the optical. The disturbed H\,{\small I} morphologies could be 
related either to the occurrence of galactic winds or to episodes of mass 
accretion and/or ingestion of low-mass companions (e.g., Kobulnicky \& 
Skillman 1995; Stil \& Israel 1998, 2002; Cannon et al. 2004). In some cases, 
both mechanisms are competing to determine the physical properties of the 
galaxy.

\section{NGC\,1569 and NGC\,1705}

Whatever its (unknown) cause, the strong recent star formation activity in 
NGC\,1569 triggered a galactic outflow whose signature can be observed in 
different bands (Martin et al. 2002). In NGC\,1705, the H\,{\small I} 
kinematics is quite regular (Meurer et al. 1998). Yet, its spectacular 
galactic wind (Meurer et al. 1992; Heckman et al. 2001) bears witness to the 
recent rather exceptional star formation activity (Annibali et al. 2003). In 
the following, we will concentrate on these two windy starburst dwarfs.

\begin{table}[ht]
\caption{Observational properties.}
\begin{tabular*}{\textwidth}{@{\extracolsep{\fill}}lclcl}
\sphline
\it Quantity\vrule height 14pt width 0pt depth 4pt & 
\multicolumn{2}{c}{NGC\,1569} & \multicolumn{2}{c}{NGC\,1705}\cr
 & \it Observed value &\it Ref. & \it Observed value &\it Ref.\cr\sphline
Distance & 2.2 $\pm$ 0.6 Mpc & 1 & 5.1 $\pm$ 0.6 Mpc & 6\cr
Gas mass & (1.5 $\pm$ 0.3) $\times$ 10$^8$ $M_\odot$ & 1 & 1.7 $\times$ 10$^8$ 
$M_\odot^{\phantom{a}a}$ & 7\cr
Total mass & 3.3 $\times$ 10$^8$ $M_\odot$ & 1 & 3.4 $\times$ 10$^8$ 
$M_\odot^{\phantom{a}a}$ & 7\cr
$Z$ & 0.004 & 2 & 0.004 & 8\cr
log(O/H)+12 & 8.19--8.37 & 3, 4, 5 & 8.21 $\pm$ 0.05 & 9\cr
log(N/O) & $-$1.39 $\pm$ 0.05 & 5 & $-$1.75 $\pm$ 0.06$^b$ & 9\cr\sphline
\end{tabular*}
\begin{tablenotes}
$^a$Gaseous and total masses were modified to reflect the distance used here.

$^b$The mean value is higher, $-$1.63 $\pm$ 0.07, if the anomalously low 
nitrogen abundance measured in region B4 is ignored (Lee \& Skillman 2004).

1-- Israel (1988); 2-- González Delgado et al. (1997); 3-- Calzetti et al. 
(1994); 4-- Martin (1997); 5-- Kobulnicky \& Skillman (1997); 6-- Tosi et al. 
(2001); 7-- Meurer et al. (1992); 8-- Storchi-Bergmann et al. (1994); 9-- Lee 
\& Skillman (2004).
\end{tablenotes}
\label{RomanoDtab1}
\end{table}

Despite being differently classified (as a DIG NGC\,1569 and as a BCD 
NGC\,1705) they display fairly similar global properties (except for the 
nitrogen abundance -- see Table~\ref{RomanoDtab1}). The reported chemical 
abundances refer to the present time (being measured from H\,{\small II} 
region optical lines). Therefore, they have to be compared with the final 
points of the theoretical tracks. The star formation history (SFH) and initial 
mass function (IMF) have been deduced from deep optical and near infrared 
\emph{HST} photometry by applying the synthetic colour-magnitude diagram (CMD) 
method (Tosi et al. 1991). An almost continuous star formation activity has 
been derived for both galaxies in the last $\sim$1 Gyr (Greggio et al. 1998; 
Annibali et al. 2003; Angeretti et al. 2005). The youngest burst in NGC\,1705, 
started 3 Myr ago, is still ongoing. Its rate, $\sim$0.3 $M_\odot$ yr$^{-1}$ 
according to Annibali et al. (2003), is comparable to that of the latest, 
strongest burst occurred in NGC\,1569 (Greggio et al. 1998). As back as the 
observations can go -- up to $\sim$1--2 Gyr ago for NGC\,1569 (Greggio et al. 
1998; Angeretti et al. 2005) and up to $\sim$5 Gyr ago for NGC\,1705 (Annibali 
et al. 2003) -- there is always evidence that the galaxy was forming stars at 
that time. The IMF is found to be close to the Salpeter value for both objects.

\subsection{Chemical evolution model results}

Taking advantage of the constraints put on SFH and IMF by the CMD analysis, we 
construct detailed chemical evolution models for NGC\,1569 and NGC\,1705 aimed 
at understanding how these galaxies form and evolve. We make the working 
hypothesis that each galaxy can be described by a one-zone model, a reasonable 
assumption as long as significant abundance gradients are not observed in 
these systems.

The basic equations we use in order to follow the temporal evolution of 
several chemical species in the gas are the same described in Bradamante et 
al. (1998). The stellar lifetimes are taken into account in detail, i.e., the 
instantaneous recycling approximation is relaxed. Up-to-date stellar 
nucleosynthesis is included in the model and galactic winds are assumed to 
originate when the thermal energy of the gas equates its binding energy. The 
thermal energy of the gas increases due to supernova (SN) explosions and 
stellar winds; according to Recchi et al. (2001), a higher thermalization 
efficiency is assigned to Type Ia SNe, because they explode in a warm medium, 
rarefied by previous Type II SN explosions. The binding energy of the gas 
depends on the dark matter (DM) amount and distribution; a massive 
($M_{\mathrm{dark}}/M_{\mathrm{lum}}$ = 10), diffuse 
($R_{\mathrm{eff}}/R_{\mathrm{dark}}$ = 0.1) DM halo is assumed. For late-type 
dwarfs, mostly metals are expected to be carried away in the outflow, while 
only a smaller fraction of unprocessed gas is likely to be affected (De Young 
\& Gallagher 1990; Mac Low \& Ferrara 1999; D'Ercole \& Brighenti 1999).

\begin{figure}[ht]
\includegraphics[width=\textwidth]{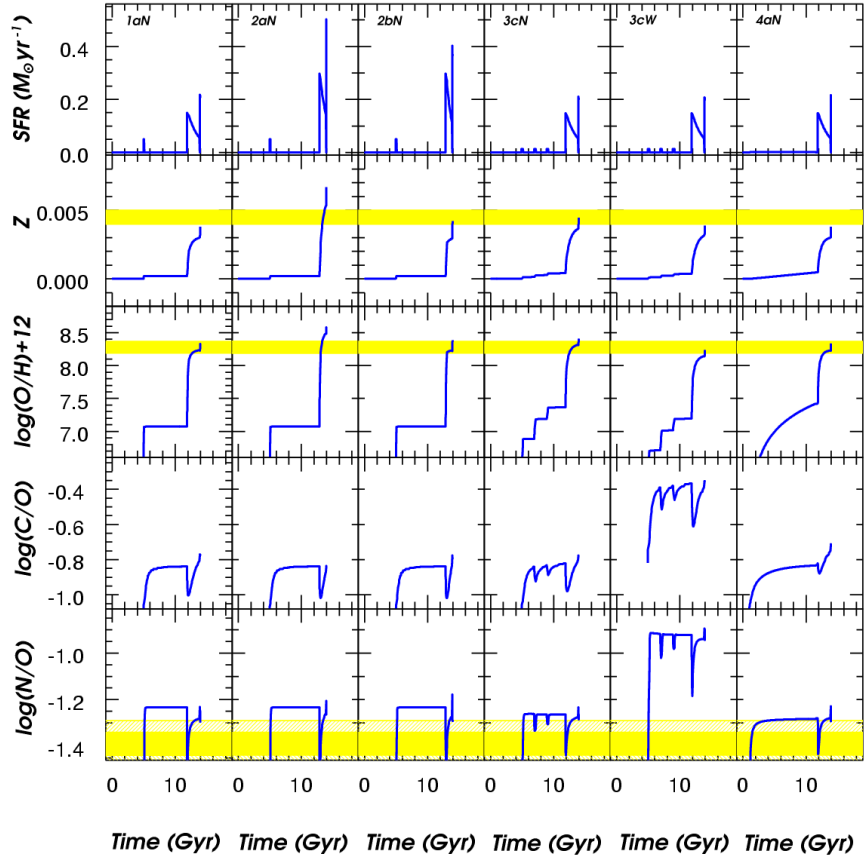}
\caption{Temporal behaviour of (i) star formation rate, (ii) metallicity, 
(iii) oxygen abundance, (iv) carbon to oxygen, and (v) nitrogen to oxygen 
abundance ratios in the gas of NGC\,1569 for six different chemical evolution 
models (see text). Predictions from different models \emph{(blue solid 
curves)} are displayed together with the observational values when available 
\emph{(yellow horizontal bands)}. The star formation rate is actually an input 
quantity for the models. Notice that measured abundances refer to H\,{\small 
II} region composition and must then be compared with the \emph{end points} of 
the theoretical tracks.}
\label{RomanoDfig1}
\end{figure}

Fig.~\ref{RomanoDfig1} displays some results obtained with different models 
for NGC\,1569. The star formation rate, metallicity, oxygen abundance, and 
carbon (nitrogen) to oxygen abundance ratios in the gas are shown as a 
function of time for six models differing in the adopted SFH at epochs where 
no constraints are available from \emph{HST} photometry, galactic wind 
efficiency and stellar nucleosynthesis. In particular, Models~\emph{1aN}, 
\emph{2aN} and \emph{2bN} assume that only one small burst of star formation 
preceded the strong last Gyr activity detected with \emph{HST}. For Models 
\emph{3cN} and \emph{3cW}, the most ancient activity consists of three weak 
short-lasting bursts, while Model~\emph{4aN} adopts a continuous, low-level 
star formation in the past. The metal ejection efficiency due to galactic 
winds is higher for Model~\emph{2bN} than for Models~\emph{1aN}, \emph{2aN} 
and \emph{4aN}, while it is slightly lower for Models~\emph{3cN} and 
\emph{3cW}. All the models share the same stellar nucleosynthesis 
prescriptions (van den Hoek \& Groenewegen 1997 yields with constant mass loss 
parameter along the AGB for low- and intermediate-mass stars, plus Nomoto et 
al. 1997 yields for massive stars), but Model~\emph{3cW} (van den Hoek \& 
Groenewegen 1997 yields with metallicity-dependent mass loss parameter along 
the AGB for low- and intermediate-mass stars, plus Woosley \& Weaver 1995 
yields for massive stars).

Detailed discussion of the model results will be presented elsewhere, here we 
limit ourselves to some basic considerations. The first thing to notice is the 
high degree of uncertainty which affects model predictions due to 
uncertainties in the stellar nucleosynthesis. By comparing the fourth and 
fifth columns of Fig.~\ref{RomanoDfig1}, it can be seen that the theoretical 
C/O and N/O may vary by $\sim$0.3--0.4 dex, depending on the assumed stellar 
yields. In particular, the set with the yields by van den Hoek \& Groenewegen 
(1997; metallicity-dependent mass loss parameter along the AGB) plus Woosley 
\& Weaver (1995) overestimates the nitrogen abundance actually seen in 
NGC\,1569. Part of the discrepancy is likely due to the fact that primary 
nitrogen production from intermediate-mass stars is overestimated by this 
yield set (see also Chiappini et al. 2003).

Another important issue is that of the galactic wind efficiency. The most 
recent, violent star formation activity in NGC\,1569 naturally triggers and 
sustains an outflow on a galactic scale in our model. The more efficient the 
star formation process, the more effective must be the wind in removing the 
newly produced metals in order to explain the H\,{\small II} region data. For 
instance, Model~\emph{2aN} (Fig.~\ref{RomanoDfig1}, second column), computed 
by assuming a star formation efficiency higher than Model~\emph{1aN} 
(Fig.~\ref{RomanoDfig1}, first column) during the last Gyr, overproduces the 
present-day oxygen content and overall metallicity\footnote{A smaller effect 
is seen in the C/O and N/O ratios, since carbon, nitrogen and oxygen are 
always ejected in the same proportions in the outflow for all the models.} of 
the galaxy, unless the efficiency of gas removal from the galaxy is increased 
(Model~\emph{2bN}, Fig.~\ref{RomanoDfig1}, third column).

Finally, it is worth stressing that, besides the standard bursting mode of 
star formation (strong bursts of star formation alternating long quiescent 
phases) often attributed to DIGs and BCDs by chemical evolution modellers 
(e.g. Bradamante et al. 1998 and references therein), assuming a continuous 
low-level star formation rate in the past also produces results in good 
agreement with the observations (Model \emph{4aN}, Fig.~\ref{RomanoDfig1}, 
last column). Because of the small differences in the model predictions in the 
two cases, it is unlikely that future measurements such as abundance 
determinations for single stars tracing the whole chemical enrichment history 
of the galaxy will allow us to discriminate between the two opposite 
scenarios. However, it should be noted that the observational evidence points 
against long periods of quiescence between successive bursts, at least for the 
look-back times actually surveyed by the observations.

\begin{figure}[ht]
\sidebyside
{\includegraphics[width=5.5cm]{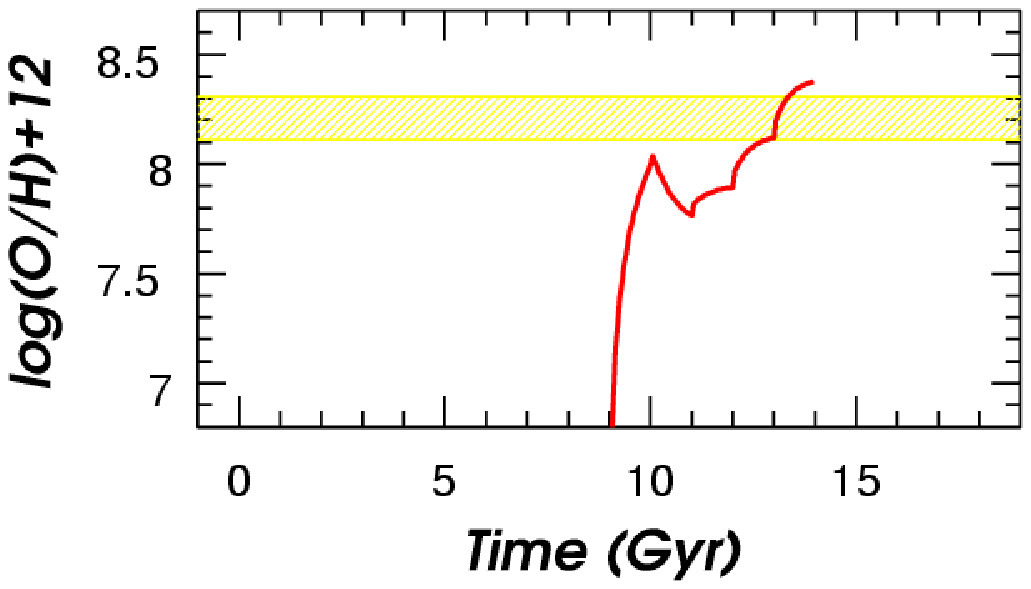}
\letteredcaption{a}{The temporal evolution of the oxygen abundance in the 
interstellar medium of NGC\,1705 predicted by the model \emph{(red solid 
line)} is compared to the available observations (\emph{yellow band;} mean 
value from [O\,{\small III}] $\lambda$\,4363 measurements by Lee \& Skillman 
2004).}
\label{RomanoDfig2a}}
{\includegraphics[width=5.5cm]{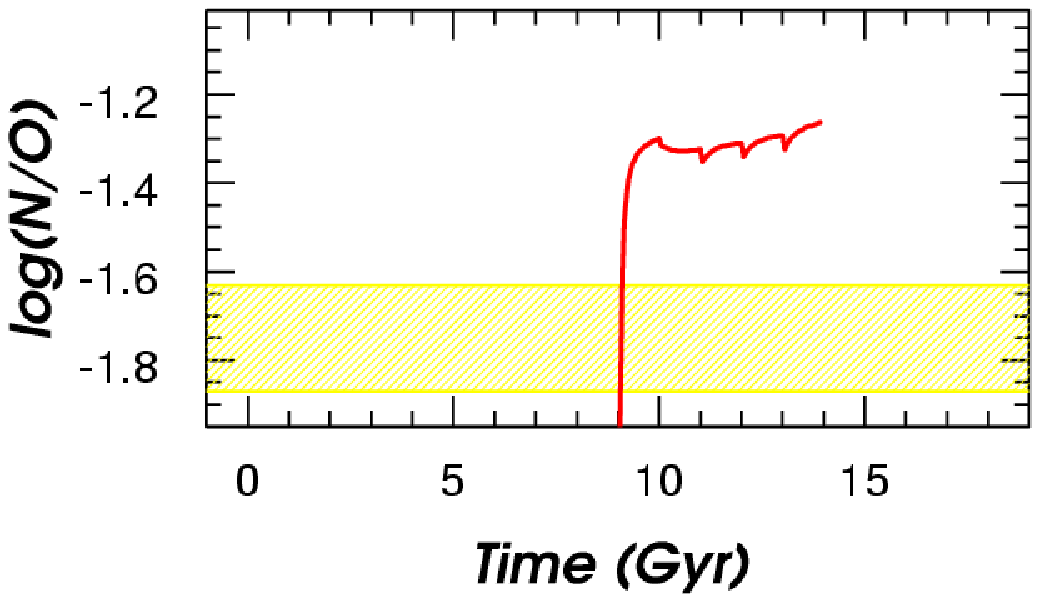}
\letteredcaption{b}{The temporal evolution of the nitrogen to oxygen abundance 
ratio in the interstellar medium of NGC\,1705 predicted by the model 
\emph{(red solid line)} is compared to the available observations (at 
2\,$\sigma$ -- \emph{yellow band;} Lee \& Skillman 2004).}
\label{RomanoDfig2b}}
\end{figure}

Models for NGC\,1705 are similarly constructed by adopting the SFH and IMF 
inferred from the observations (Annibali et al. 2003). The stellar yields, 
galactic wind onset conditions and efficiency are the same adopted by 
successful models for NGC\,1569. While the total and gaseous mass of the 
system at the present time as well as its current metallicity and oxygen 
content are easily reproduced, the predicted nitrogen abundance turns out to 
be always higher than measured from H\,{\small II} regions (cfr. 
Figs.~\ref{RomanoDfig2a} and \ref{RomanoDfig2b}).

\subsubsection{Interpreting the N/O data}

Understanding the origin of nitrogen is one of the major goal of modern 
astrophysics. Since the pioneering work of Edmunds \& Pagel (1978), a stellar 
source of primary N has became strongly in demand. Recent abundance data for 
very metal-poor Galactic halo stars (Spite et al. 2004; Israelian et al. 2004) 
suggest that an important production of primary N actually took place in the 
first generation of massive halo stars, while delayed primary N production 
from intermediate-mass stars likely overwhelms any massive star contribution 
at later times (e.g. Chiappini et al. 2003). However, current stellar yields 
probably overestimate the amount of primary nitrogen produced through hot 
bottom burning in intermediate-mass stars. Clearly, changing the 
nucleosynthesis prescriptions in such a way that the N/O ratio measured for 
NGC\,1705 is reproduced, immediately destroys the agreement between model 
predictions and observations for NGC\,1569, since a too low N/O ratio is 
obtained in this case. On the other hand, modifying the intermediate- and 
high-mass star stellar mass spectrum within the range allowed by \emph{HST} 
observations does not offer a viable solution, either. One is left with the 
possibility that the relative fractions of nitrogen and oxygen lost from the 
two galaxies are different. Alternatively, in NGC\,1705 we might be seing 
localized self-pollution due to dying young massive stars born during the last 
3 Myr of star formation activity. Both such hypotheses wait for detailed 
hydrodynamical computations in order to be verified.







\notes

\begin{acknowledgments}
DR wish to thank A. Aloisi, H. Lee and E. Skillman for discussions.
\end{acknowledgments}

\end{document}